\documentclass[aip,jcp,reprint,groupedaddress,floatfix]{revtex4-2}
\pdfoutput=1

\usepackage{graphicx,amsmath}
\usepackage{hyperref}
\hypersetup{colorlinks=true, citecolor=blue, urlcolor=blue, linkcolor=blue}
\usepackage{xcolor}
\usepackage{siunitx}

\newcommand{\kB}{k_\mathrm{B}}
\newcommand{\etal}{{\textit{et al.}}}
\newcommand{\Rh}{R_\mathrm{h}}

\begin{document}

% Use the \preprint command to place your local institutional report
% number in the upper righthand corner of the title page in preprint mode.
% Multiple \preprint commands are allowed.
% Use the 'preprintnumbers' class option to override journal defaults
% to display numbers if necessary
%\preprint{}

%Title of paper
\title{Isotope effect on the anomalies of water: a corresponding states analysis}

% repeat the \author .. \affiliation  etc. as needed
% \email, \thanks, \homepage, \altaffiliation all apply to the current
% author. Explanatory text should go in the []'s, actual e-mail
% address or url should go in the {}'s for \email and \homepage.
% Please use the appropriate macro foreach each type of information

% \affiliation command applies to all authors since the last
% \affiliation command. The \affiliation command should follow the
% other information
% \affiliation can be followed by \email, \homepage, \thanks as well.
\author{Fr\'ed\'eric Caupin}
\email[]{frederic.caupin@univ-lyon1.fr}
\author{Pierre Ragueneau}
\author{Bruno Issenmann}
\email[]{bruno.issenmann@univ-lyon1.fr}
%\homepage[]{Your web page}
%\thanks{}
%\altaffiliation{}
\affiliation{Institut Lumi\`ere Mati\`ere, Universit\'e Claude Bernard Lyon 1, CNRS, Institut Universitaire de France, F-69622, Villeurbanne, France}
%Collaboration name if desired (requires use of superscriptaddress
%option in \documentclass). \noaffiliation is required (may also be
%used with the \author command).
%\collaboration can be followed by \email, \homepage, \thanks as well.
%\collaboration{}
%\noaffiliation

\date{\today}

\begin{abstract}
%\textcolor{red}{Remplacer documentclass[aps,prx,reprint,groupedaddress,floatfix]{revtex4} par documentclass[aps,prx,reprint,groupedaddress,floatfix]{revtex4-2}}
Light and heavy water show similar anomalies in thermodynamic and dynamic properties, with a consistent trend of anomalies occurring at higher temperature in heavy water. Viscosity also increases faster upon cooling in heavy water, causing a giant isotope effect, with a viscosity ratio near 2.4 at \SI{244}{K}. While a simple temperature shift apparently helps in collapsing experimental data for both isotopes, it lacks a clear justification, changes value with the property considered, and requires additional \textit{ad hoc} scaling factors. Here we use a corresponding states analysis based on the possible existence of a liquid-liquid critical point in supercooled water. This provides a coherent framework which leads to the collapse of thermodynamic data. The ratio between dynamic properties of the isotopes is strongly reduced. In particular, the decoupling between viscosity $\eta$ and self-diffusion $D$, measured as a function of temperature $T$ by the Stokes-Einstein ratio $D\eta/T$, is found to collapse after applying the corresponding states analysis. Our results are consistent with simulations and suggest that the various isotope effects mirror the one on the liquid-liquid transition.
\end{abstract}

% insert suggested keywords - APS authors don't need to do this
%\keywords{}

%\maketitle must follow title, authors, abstract, and keywords
\maketitle

% body of paper here - Use proper section commands
% References should be done using the \cite, \ref, and \label commands

%\textbf{Introduction}

% Put \label in argument of \section for cross-referencing
%\textbf{\label{}}
%aaa

\textbf{INTRODUCTION\label{sec:intro}}

At ambient conditions, water (H$_2$O) and its fully deuterated isotope, deuterium oxide (D$_2$O), have almost identical molar volume or surface tension, but their shear viscosities differ by nearly 25\,\%. Isotopic content matters for higher living organism, high doses of heavy water being lethal, because of slowing down of chemical reaction kinetics~\cite{francl_weight_2019}. Heavy water exhibits the same anomalies that are found in light water~\cite{holten_thermodynamics_2012,gallo_water_2016,herrig_reference_2018}, e.g. density maximum, isothermal compressibility and isobaric heat capacity minima near ambient temperature, or non-monotonic pressure dependence of viscosity and diffusivity at low temperature. In both isotopes, these anomalies get more pronounced when the liquid enters the supercooled region, below the ice-liquid equilibrium temperature. Several theories have been put forward to explain these anomalies, including the existence of a metastable phase transition between two distinct supercooled liquids~\cite{poole_phase_1992}. Experiments supporting this possibility have been reported for H$_2$O: a first-order like transition between two glassy phases of water, differing in structure and density~\cite{mishima_apparently_1985}, and a discontinuity in decompression-induced melting lines of high-pressure ices~\cite{mishima_decompression-induced_1998}. The two amorphous ices~\cite{klug_raman_1986} and the melting line discontinuity~\cite{mishima_liquidliquid_2000} are also found in D$_2$O.

\begin{figure}[ttt]
\centering
\includegraphics[width=0.94\columnwidth]{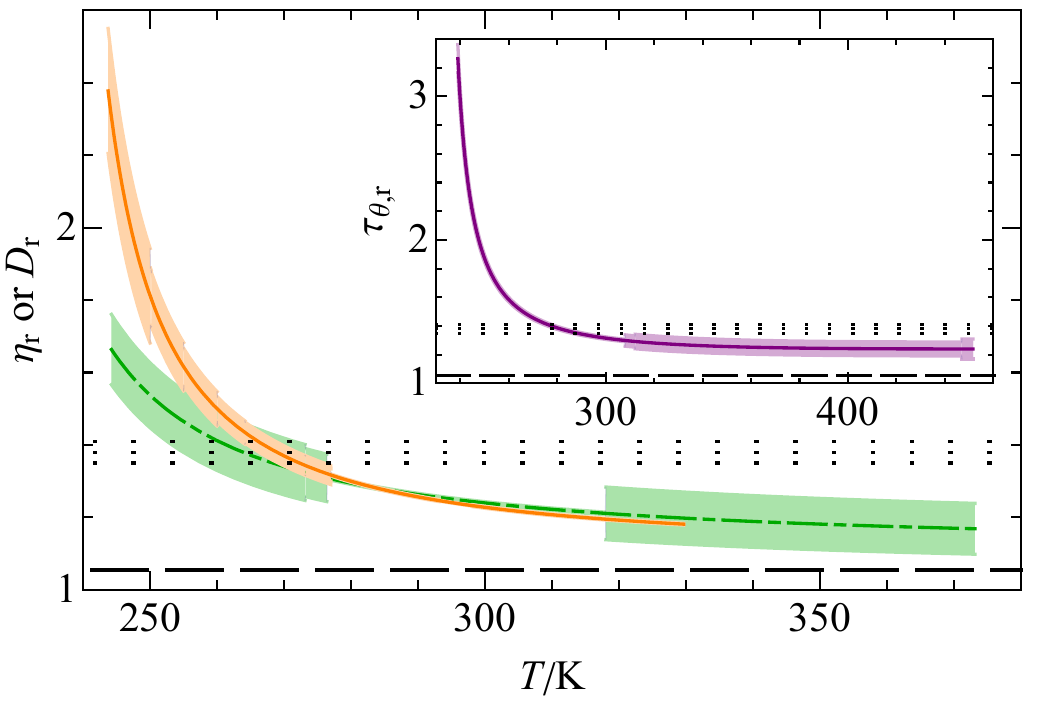}
\caption{Isotope effect for viscosity ($\eta_\mathrm{r}=\eta_\mathrm{D}/\eta_\mathrm{H}$, solid orange curve) and self-diffusion ($D_\mathrm{r}=D_\mathrm{H}/D_\mathrm{D}$, dash-dotted green curve) for water. The values of $\sqrt{M_\mathrm{r}}$ and $\sqrt{I_\mathrm{r}}$ are shown with horizontal dashed and dotted lines, respectively. The inset displays the isotope effect for the rotational correlation time ($\tau_\mathrm{r}=\tau_{\theta,\mathrm{D}}/\tau_{\theta,\mathrm{H}}$, purple curve). The colored areas denote the $1-\sigma$ uncertainties.
\label{fig:IE}
}
\end{figure}

One of the strongest contrast between H$_2$O and D$_2$O is observed for their dynamic properties. Figure~\ref{fig:IE} shows the ratios $\eta_\mathrm{r}=\eta_\mathrm{D}/\eta_\mathrm{H}$, $D_\mathrm{r}=D_\mathrm{H}/D_\mathrm{D}$, and $\tau_{\theta,\mathrm{r}}=\tau_{\theta,\mathrm{D}}/\tau_{\theta,\mathrm{H}}$, where $\eta$ is the shear viscosity, $D$ the self-diffusion coefficient, and $\tau_\theta$ the rotational correlation time, respectively, and subscripts H and D refer to the light and heavy isotope, respectively. All experimental data used in this work are presented in Tables~\ref{tab:H2O} and \ref{tab:D2O} for H$_2$O and D$_2$O, respectively. Our recent viscosity data~\cite{dehaoui_viscosity_2015,ragueneau_shear_2022} allow us to plot $\eta_\mathrm{r}$ to lower temperature than before. It reaches a massive 2.38 at \SI{243.7}{K}, which, to our knowledge, is only exceeded by quantum liquids: helium 4, which becomes superfluid at higher temperature than helium 3; and the hydrogen isotopes, H$_2$ and D$_2$, for which $\eta_\mathrm{r} \simeq 3$~\cite{rudenko_viscosity_1963}. $\tau_{\theta,\mathrm{r}}$ reaches an even higher value, 3.26 at \SI{239}{K}.

To reconcile density and viscosity data for light and heavy water, Robinson and collaborators introduced the thermal offset hypothesis~\cite{vedamuthu_properties_1994,vedamuthu_simple_1996,cho_thermal_1999}. We first present this approach and its limitations. Then, we attempt another analysis, based on the idea of corresponding states\cite{guggenheim_principle_1945,su_modified_1946,kostrowickawyczalkowska_thermodynamic_2000,abdulkadirova_thermodynamic_2002}. Assuming the existence of a liquid-liquid critical point terminating a metastable liquid-liquid transition in supercooled water~\cite{gallo_water_2016}, we show how a simple rescaling of temperature and pressure by the critical values leads to a striking collapse of thermodynamic and dynamic data for the two isotopes. We finally discuss the connection of our corresponding states analysis with the thermal offset hypothesis and with other works on the liquid-liquid transition in water.

\begin{table}[ttt]
\caption{Properties of light water used in this study. The last column indicates the type of fitting function used in the calculations for comparison with heavy water data.\label{tab:H2O}}
\begin{ruledtabular}
\begin{tabular}{cccc}
Ref. & $P$ (MPa)	& $T$ range (K) & Fitting function\\
\hline
\multicolumn{4}{c}{Molar volume $V_\mathrm{mol}$}\\
\onlinecite{hare_density_1987}	& 0.1 & $239.74-267.92$	& 8-th order \\
\onlinecite{theinternationalassociationforthepropertiesofwaterandsteam_revised_2018}\footnotemark[1]	& 0.1 & $273.15-313.15$	&  polynomial\\
\hline
\multicolumn{4}{c}{Isothermal compressibility $\kappa_T$}\\
\onlinecite{kanno_water_1979} & $10$ & $250.35-297.95$ & 4-th order polynomial\\
\onlinecite{kanno_water_1979} & $50$ & $248.15-298.15$ & 4-th order polynomial\\
\onlinecite{kanno_water_1979} & $100$ & $240.95-298.05$ & 4-th order polynomial\\
\hline
\multicolumn{4}{c}{Isobaric heat capacity $C_P$}\\
\onlinecite{angell_heat_1982} & 0.1 & $236.01-290$ & 6-th order polynomial\\
\hline
\multicolumn{4}{c}{Viscosity $\eta$}\\
\onlinecite{ragueneau_shear_2022}\footnotemark[2] & 0.1 & $239.15-491.95$ & Speedy-Angell law\footnotemark[4]	\\
\hline
\multicolumn{4}{c}{Self-diffusion coefficient $D$}\\
\onlinecite{dehaoui_viscosity_2015}\footnotemark[3] & 0.1 & $237.8-498.2$ & Speedy-Angell law\footnotemark[4]	\\
%\onlinecite{prielmeier_pressure_1988} & $50$ & $243-273$	& 8 & 	\\
%\onlinecite{prielmeier_pressure_1988} & $100$ & $238-273$	& 9 &  power law	\\
%\onlinecite{prielmeier_pressure_1988} & $150$ & $228-273$	& 11 &  	\\
%\onlinecite{prielmeier_pressure_1988} & $200$ & $218-273$	& 13 & 	\\
\hline
\multicolumn{4}{c}{Rotational correlation time $\tau_{\theta}$}\\
\onlinecite{dehaoui_viscosity_2015}\footnotemark[3] & 0.1 & $236.2-451.6$ & Speedy-Angell law\footnotemark[4]	\\
\end{tabular}
\footnotetext[1]{We used this formulation to compute $V_\mathrm{mol}$ values every \SI{5}{K}.}
\footnotetext[2]{This reference compiles data from a series of sources~\cite{hallett_temperature_1963,korosi_density_1981,eicher_high-precision_1971,kestin_free_1981,collings_high_1983,kestin_viscosity_1985,berstad_accurate_1988,dehaoui_viscosity_2015}.}\footnotetext[3]{This reference compiles data from a series of sources for $D$~\cite{mills_self-diffusion_1973,krynicki_pressure_1978,easteal_diaphragm_1989,price_selfdiffusion_1999}
and for $\tau_{\theta}$~\cite{hindman_relaxation_1974,qvist_rotational_2012}.}
\footnotetext[4]{$X(T)=X_0 (T/T_\mathrm{s} - 1)^{-\gamma}$, with parameters $X_0$, $T_\mathrm{s}$, and $\gamma$ given in Ref.~\onlinecite{ragueneau_shear_2022} for $X=\eta$, and in Ref.~\onlinecite{dehaoui_viscosity_2015} for $X=D$ and $\tau_{\theta}$. The fits are valid up to \SI{348.15}{K}, \SI{498.2}{K}, and \SI{451.63}{K} for $\eta$, $D$, and $\tau_{\theta}$, respectively.}
\end{ruledtabular}
\end{table}

\begin{table}[ttt]
\caption{Properties of heavy water used in this study. \label{tab:D2O}}
\begin{ruledtabular}
\begin{tabular}{ccc}
Ref. & $P$ (MPa)	& $T$ range (K) \\
\hline
\multicolumn{3}{c}{Molar volume $V_\mathrm{mol}$}\\
\onlinecite{holten_thermodynamics_2012}\footnotemark[1]	& 0.1 & $244.1-313.14$\\
\onlinecite{blahut_relative_2019}\footnotemark[2]	& 0.1 & $258.15-298.15$\\
\hline
\multicolumn{3}{c}{Isothermal compressibility $\kappa_T$}\\
\onlinecite{kanno_water_1979} & $10$ & $253.36-297.8$ \\
\onlinecite{kanno_water_1979} & $50$ & $248.9-298$\\
\onlinecite{kanno_water_1979} & $100$ & $244.2-297.9$\\
\hline
\multicolumn{3}{c}{Isobaric heat capacity $C_P$}\\
\onlinecite{angell_heat_1982} & 0.1 & $236.01-290$\\
\hline
\multicolumn{3}{c}{Viscosity $\eta$}\\
\onlinecite{ragueneau_shear_2022}\footnotemark[3] & 0.1 & $243.7-493.05$\\
\hline
\multicolumn{3}{c}{Self-diffusion coefficient $D$}\\
\onlinecite{ragueneau_shear_2022}\footnotemark[3] & 0.1 & $244.2-623$\\
%\onlinecite{prielmeier_pressure_1988} & $50$ & $243-273$	& 8\\
%\onlinecite{prielmeier_pressure_1988} & $100$ & $238-273$	& 9\\
%\onlinecite{prielmeier_pressure_1988} & $150$ & $228-273$	& 11\\
%\onlinecite{prielmeier_pressure_1988} & $200$ & $218-273$	& 13\\
\hline
\multicolumn{3}{c}{Rotational correlation time $\tau_{\theta}$}\\
\onlinecite{ragueneau_shear_2022}\footnotemark[3] & 0.1 & $239.0-473.15$\\
\end{tabular}
\end{ruledtabular}
\footnotetext[1]{This reference compiles data from a series of sources~\cite{kell_precise_1967,zheleznyi_density_1969,rasmussen_clustering_1973,emmet_specific_1975,kanno_volumetric_1980,hare_densities_1986}.}
\footnotetext[2]{Values from Table~V of Ref.~\onlinecite{blahut_relative_2019}.}
\footnotetext[3]{This reference compiles data from a series of sources for $\eta$~\cite{hardy_viscosity_1949,selecki_viscosity_1970,millero_density_1971,goncalves_viscosity_1980,agayev_experimental_1980,kestin_viscosity_1985,agayev_heavywater_1990}, $D$~\cite{mills_self-diffusion_1973,woolf_tracer_1976,weingartner_diffusion_1984,prielmeier_pressure_1988,price_temperature_2000,hardy_isotope_2001,yoshida_selfdiffusion_2008}, and $\tau_{\theta}$~\cite{hindman_relaxation_1971,jonas_molecular_1976,lang_pressure_1980,matubayasi_structural_2001,ropp_rotational_2001,hardy_isotope_2001,qvist_rotational_2012}.}
\end{table}

\textbf{THE THERMAL OFFSET HYPOTHESIS\label{sec:Tshift}}

\textit{The thermal offset concept.\label{sec:Tshiftdef}} In a series of works~\cite{vedamuthu_properties_1994,vedamuthu_simple_1996,cho_thermal_1999}, Robinson and collaborators proposed that several properties of D$_2$O can be deduced from those of H$_2$O by a simple relation:
\begin{equation}
X_\mathrm{D} (T) = \lambda X_\mathrm{H} (T - \Delta T) \, \label{eq:Tshift}
\end{equation}
where $X_\mathrm{D}$ and $X_\mathrm{H}$ are the values of a property for D$_2$O and H$_2$O, respectively, $\lambda$ is an amplitude factor, and $\Delta T$ is a thermal offset.

\begin{figure}
\centering
\includegraphics[width=0.94\columnwidth]{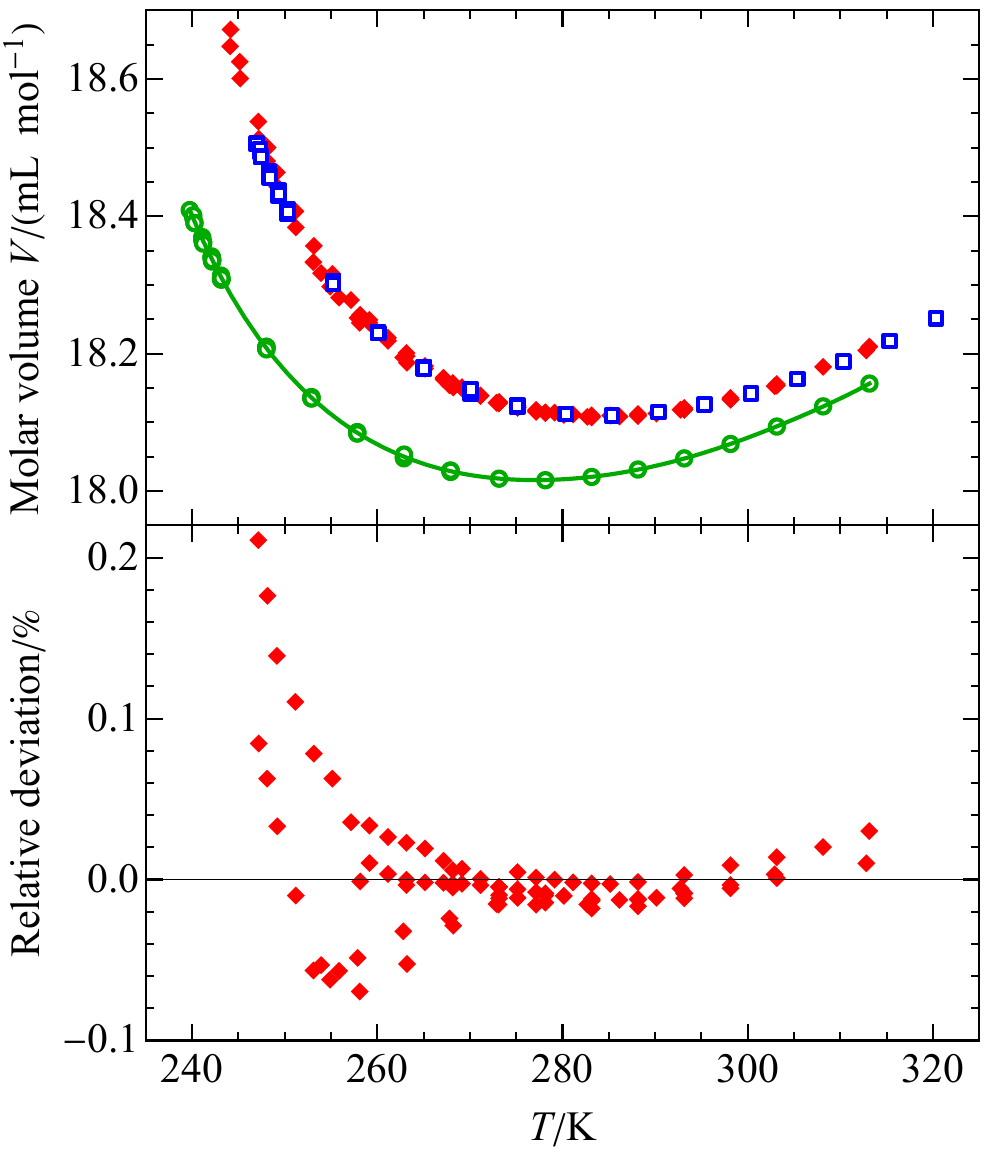}
\caption{Molar volume of water. Top panel: data for H$_2$O (empty green circles) with their fit (green curve), D$_2$O (filled red diamonds), and results of Eq.~\ref{eq:Tshift} applied to H$_2$O data (empty blue squares). Bottom panel: relative deviation of D$_2$O data from values calculated using Eq.~\ref{eq:Tshift} combined with the fit to H$_2$O data. Here $\Delta T=\SI{7.2}{K}$ and $\lambda=M_\mathrm{r}/1.1059$.\label{fig:Vmolshift}}
\end{figure}

They first applied this concept to density~\cite{vedamuthu_simple_1996}, and obtained excellent results at ambient pressure from 243 to \SI{303}{K}, with $\Delta T  = \SI{7.2}{K}$,  and $\lambda = 1.1059$. In this case, $\Delta T$ is obviously the difference between the temperatures of density maximum for the two isotopes. The factor $\lambda$ is slightly less than the ratio of their molecular weights, $M_\mathrm{r}=20.02292/18.010565 = 1.11173$, which is attributed to slightly different hydrogen-bond distances and atomic root-mean-square displacements. In Fig.~\ref{fig:Vmolshift}, we illustrate the results on molar volumes rather than density to more directly show the isotopic difference.

\begin{figure}
\centering
\includegraphics[width=0.94\columnwidth]{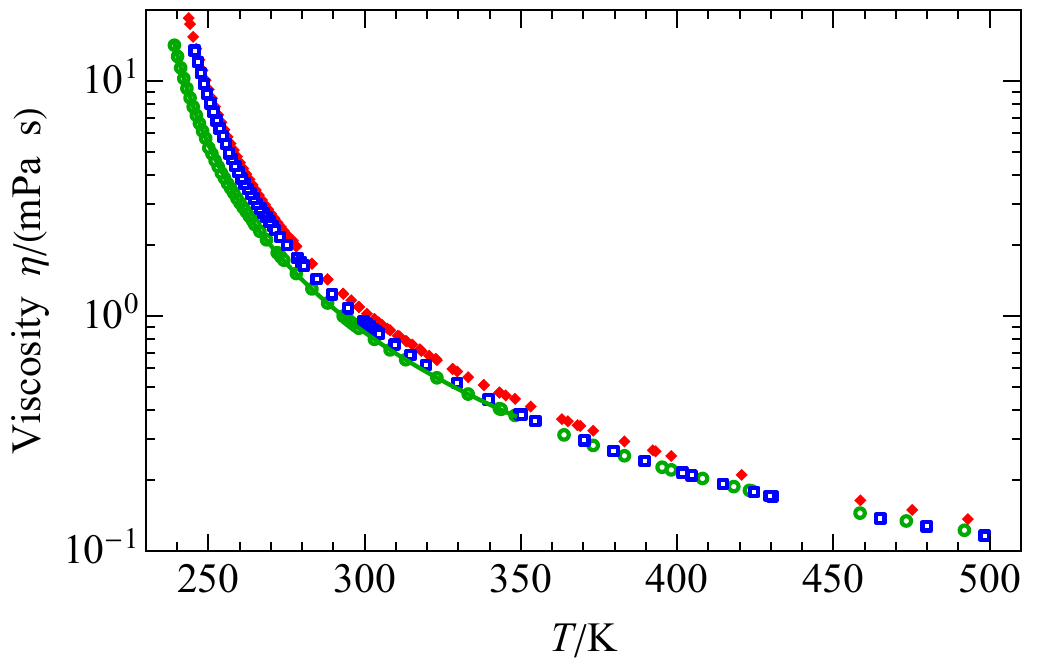}\\[1mm]
\includegraphics[width=0.94\columnwidth]{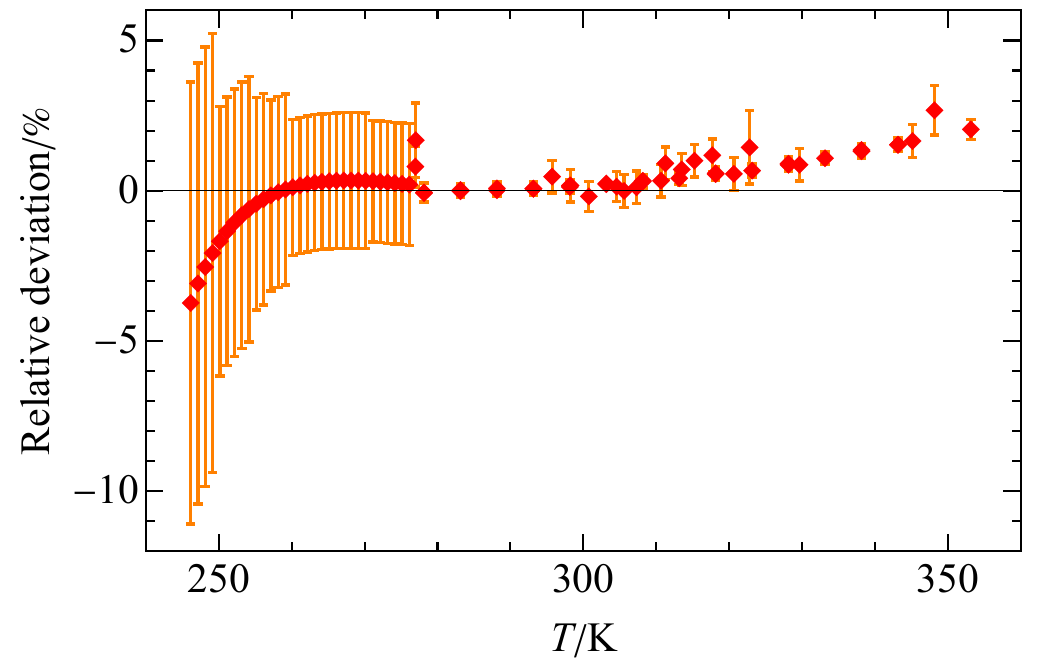}
\caption{Shear viscosity of water. Top panel: data for H$_2$O (empty green circles) with their fit (green curve), for D$_2$O (filled red diamonds), and results of Eq.~\ref{eq:Tshift} applied to H$_2$O data (empty blue squares). Bottom panel: relative deviation of D$_2$O data from values calculated using Eq.~\ref{eq:Tshift} combined with the fit to H$_2$O data. Here $\Delta T=\SI{6.498}{K}$ and $\lambda=\sqrt{M_\mathrm{r}}$. The error bars indicate one standard deviation.\label{fig:etashift}}
\end{figure}

In Ref.~\onlinecite{vedamuthu_simple_1996}, they briefly mentioned viscosity, which they investigated in details in a subsequent paper~\cite{cho_thermal_1999}. They could reproduce experimental data at ambient pressure to better than 1\% below \SI{323}{K}, using this time $\Delta T=\SI{6.498}{K}$ and $\lambda = \sqrt{M_\mathrm{r}}$. This latter value is consistent with the prediction of the gas kinetic Chapman-Enskog theory for hard spheres~\cite{chapman_mathematical_1990}. As shown in Fig.~\ref{fig:etashift}, the deviation in fact exceeds 3\% at low temperatures. This comes from the fact that, in that range, we used our low temperature data on H$_2$O and D$_2$O~\cite{ragueneau_shear_2022}, instead of the extrapolated values used in Ref.~\onlinecite{cho_thermal_1999} between 243.15 and \SI{268.15}{K}. Nevertheless, in view of the experimental uncertainties, the collapse remains satisfactory.

Harris pursued the effort, analyzing viscosity and self-diffusion data for water under pressure~\cite{harris_isotope_2002}. He concluded that the thermal offset hypothesis (Eq.~\ref{eq:Tshift}) with $\Delta T=\SI{6.498}{K}$ holds within experimental uncertainty below $\SI{303}{K}$ for viscosity up to $\SI{900}{MPa}$ with $\lambda = \sqrt{M_\mathrm{r}}$, and for self-diffusion up to $\SI{400}{MPa}$ with $\lambda = 1/\sqrt{M_\mathrm{r}}$. The Chapman-Enskog prediction is thus again retrieved.

\textit{Limitations.\label{sec:Tshiftlim}} Despite its apparent success, the thermal offset approach suffers from two major limitations.

First, the origin of the temperature offset itself is not fully clear. Robinson and co-workers provide a justification based on a mixture model for water, in which the properties of water are obtained by a simple weighted average of two components~\cite{vedamuthu_properties_1994,vedamuthu_simple_1996}. The fractions $f$ and $1-f$ of the two components are temperature dependent, but this dependence would simply be shifted in temperature between isotopes, because of ``zero-point effects in the temperature-dependent intermolecular potentials, particularly those related to molecular rotational librations.''~\cite{vedamuthu_properties_1994} The mixture model used in this approach is rather crude, being that of an ideal mixture, whereas it is thought that non-ideality plays an important role~\cite{anisimov_thermodynamics_2018,caupin_thermodynamics_2019}. It is not clear neither why the invoked zero-point effects would just cause a simple shift in temperature. Moreover, the required offset temperature $\Delta T$ differs between the density data on the one hand, and the dynamic data on the other hand.

Another difficulty lies in the statement that the Chapman-Enskog law is expected to hold in water, after the thermal offset has been removed. The viscosity ratio is indeed $\sqrt{M_\mathrm{r}}$ for isotopes of dilute monatomic gases, but the transposition to dense molecular liquids is not straightforward. Coupling between translation and rotation, possibly temperature dependent, renders the picture rather complex.

For a series of standard solvents near room temperature, Holz~\etal~\cite{holz_experimental_1996} found $\eta_\mathrm{r}$ and $D_\mathrm{r}$ to be close to each other, and close to $\sqrt{I_\mathrm{r}}$ (where $I_\mathrm{r}=I_2/I_1$ is the ratio of moments of inertia for the isotopes), rather than to $\sqrt{M_\mathrm{r}}$. Holz~{\etal} attributed this effect to a strong translation-rotation coupling. This was criticized by Buchhauser~\etal~\cite{buchhauser_selfdiffusion_1999}, who noted that up to three different principal moment of inertia may be defined for a molecule, and pointed out several cases (including water) where $D_\mathrm{r}$ varies noticeably with temperature, whereas $I_\mathrm{r}$ is constant. Figure~\ref{fig:IE} shows that, for water, $\eta_\mathrm{r}$ and $D_\mathrm{r}$ are always far above $\sqrt{M_\mathrm{r}}$, and below $260\,\mathrm{K}$, they both exceed the highest $\sqrt{I_\mathrm{r}}$. Figure~\ref{fig:IE} also shows that, while at temperatures above melting $\eta_\mathrm{r}$ and $D_\mathrm{r}$ track each other, they start departing strongly in the supercooled region.

We think it should not be a prerequisite that the dynamic isotope effects would be temperature-independent and equal to a specific value, such as $\sqrt{M_\mathrm{r}}$ or $\sqrt{I_\mathrm{r}}$. In fact, Cho~\textit{et al.} themselves~\cite{cho_thermal_1999}, noting the increasing discrepancy with experimental viscosity data at higher temperature, mentioned that an exact agreement ``can be obtained by empirically adjusting'' $\Delta T$ and $\lambda$ ``to give these parameters a temperature dependence''.

\textbf{CORRESPONDING STATES ANALYSIS\label{sec:CSA}}

\textit{Working hypothesis.\label{sec:hypo}} We propose here to improve over the thermal offset concept using a corresponding states analysis. Our goal is to provide a physically based explanation for the connection between various thermodynamic and dynamic properties of water isotopes.

Our working hypothesis is that water anomalies are due to a first order liquid-liquid transition (LLT) in the supercooled region. The liquid-liquid critical point (LLCP) terminating this LLT influences the behavior of liquid water in the supercritical region. Since it was first proposed by simulations with the ST2 potential~\cite{poole_phase_1992}, this hypothesis has increasingly gained support from simulations and experiments~\cite{gallo_water_2016}. The existence of a LLT in simulations has been firmly established by state-of-the-art free energy calculations for the ST2 water model~\cite{palmer_metastable_2014}, and evidence for a LLCP has been found with the TIP4P/2005 and TIP4P/ice water models~\cite{debenedetti_second_2020}. Transient observation of liquid-liquid coexistence in H$_2$O has been reported experimentally~\cite{kim_experimental_2020}.

The corresponding states analysis consists in writing the equation of state of water in reduced temperature -pressure ($T-P$) coordinates $\hat{T}=T/T_\mathrm{c}$ and $\hat{P}=P/P_\mathrm{c}$:
\begin{equation}
%V(T,P) = \frac{R T_\mathrm{c}}{P_\mathrm{c}} \,\hat{V}(\hat{T},\hat{P}) \;\mathrm{where} \,\hat{T}=\frac{T}{T_\mathrm{c}} \; \mathrm{and} \;\hat{P}=\frac{P}{P_\mathrm{c}} \, \label{eq:CSA}
V(T,P) = \frac{R T_\mathrm{c}}{P_\mathrm{c}} \,\hat{V}(\hat{T},\hat{P}) \, , \label{eq:CSA}
\end{equation}
with $V$ the molar volume, $R$ the gas constant, $T_\mathrm{c}$ and $P_\mathrm{c}$ the temperature and pressure of the LLCP, respectively, and $\hat{V}$ a non-dimensional universal function. Note that Eq.~\ref{eq:CSA} corresponds to the modified corresponding states principle first proposed by Su~\cite{su_modified_1946}. The quantity $R T_\mathrm{c}/P_\mathrm{c}$ is used as the volume scale, instead of the critical molar volume $V_\mathrm{c}$. The two choices are equivalent for fluids with equal critical compressibility factors $Z_\mathrm{c}=P_\mathrm{c} V_\mathrm{c}/(R T_\mathrm{c})$. However, in the case of the liquid-vapor critical point, Eq.~\ref{eq:CSA} performs better in giving a universal description of many fluids with various $Z_\mathrm{c}$~\cite{su_modified_1946}.

We thus assume that the main isotope effect is to change ($T_\mathrm{c}$,$P_\mathrm{c}$), but not the function $\hat{V}$, allowing a mapping of properties of light and heavy water. The mapping should work best at low temperature and in the supercooled liquid region, where the influence of the LLCP is stronger. We will investigate how this hypothesis applies to experimental data in the following.

\textit{Molar volume.\label{sec:VmolCSA}} As in Ref.~\onlinecite{vedamuthu_simple_1996}, we first consider the molar volumes at ambient pressure, $V_\mathrm{H}$ and $V_\mathrm{D}$ for light and heavy water, respectively. Because $P_\mathrm{c}$ is estimated to exceed \SI{50}{MPa} (see Discussion), we assume $\hat{P}\simeq 0$ at ambient pressure. Eq.~\ref{eq:CSA} then predicts the following relation:
\begin{equation}
V_\mathrm{D}(T) = \frac{\Theta}{\Pi} \,V_\mathrm{H}\!\left( \frac{T}{\Theta} \right) \,\mathrm{with}\, \Theta = \frac{T_\mathrm{c,D}}{T_\mathrm{c,H}} \, \mathrm{and}\, \Pi = \frac{P_\mathrm{c,D}}{P_\mathrm{c,H}}\, .\label{eq:rescaleV}
\end{equation}

To obtain $\Theta$ and $\Pi$, we proceed as follows. We first take molar volumes of light water at $P=\SI{0.1}{MPa}$ from the data listed in Table~\ref{tab:H2O}. We fit them with a 8-th order polynomial in $T$, which reproduces the data within their uncertainties. For heavy water, we take molar volumes at $P=\SI{0.1}{MPa}$ from the data listed in Table~\ref{tab:D2O}. Finally, we make a least-squares fit with Eq.~\ref{eq:rescaleV}, using the 8-th order polynomial for $V_\mathrm{H}$. We progressively include D$_2$O data points at lower temperatures, until there is no supporting H$_2$O data available at the corresponding temperature $T/\Theta$, which excludes 4 points below $\SI{246}{K}$. This procedure yields $\Theta=1.031$ and $\Pi=1.026$.

\begin{figure}
\centering
\includegraphics[width=0.94\columnwidth]{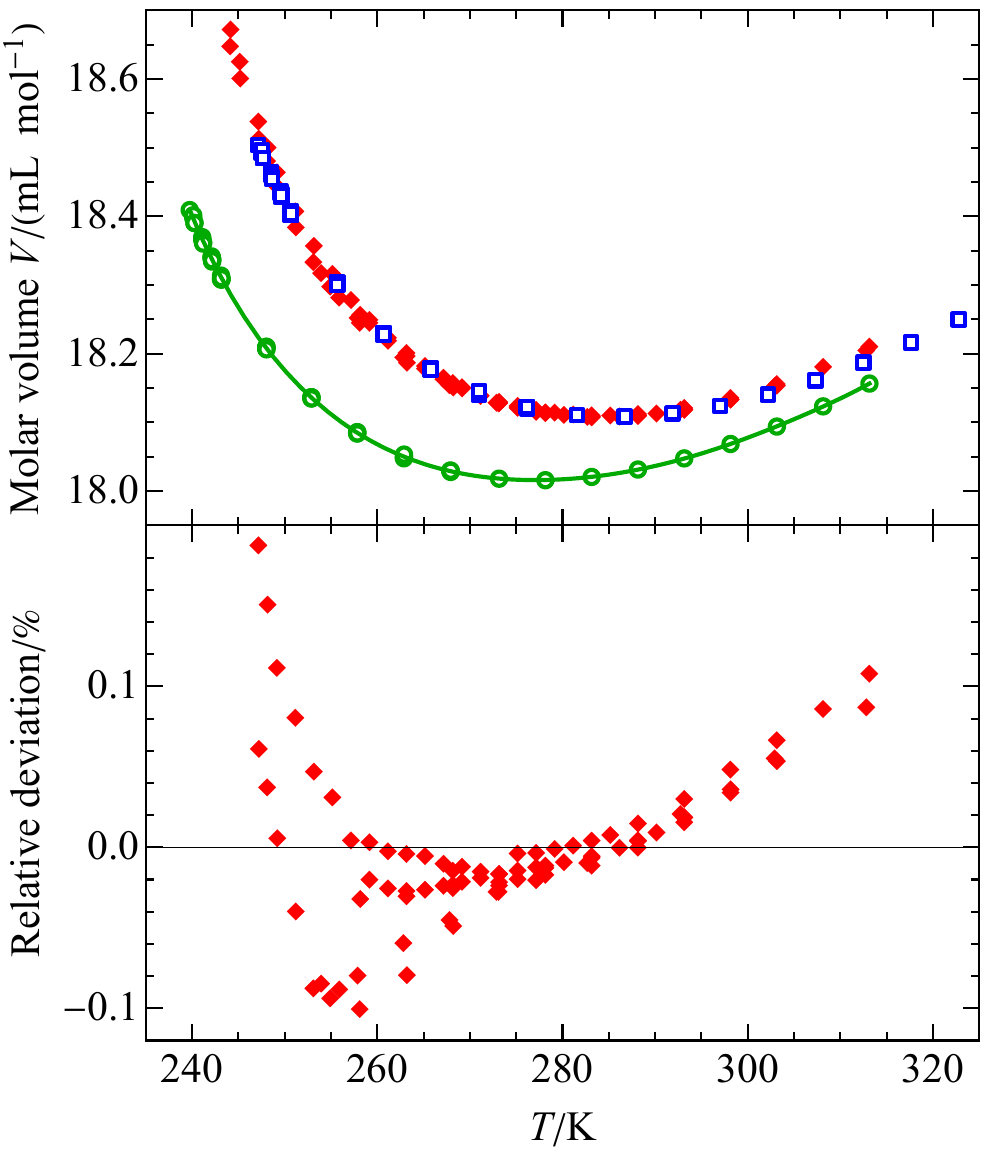}
\caption{Molar volume of water. Top panel: data for H$_2$O (empty green circles) with their fit (green curve), for D$_2$O (filled red diamonds), and results of Eq.~\ref{eq:rescaleV} applied to H$_2$O data (empty blue squares). Bottom panel: relative deviation of D$_2$O data from values calculated using Eq.~\ref{eq:rescaleV} combined with the fit to H$_2$O data. Here $\Theta=1.031$ and $\Pi=1.026$.\label{fig:VmolCSA}}
\end{figure}

Figure~\ref{fig:VmolCSA} shows the comparison between experimental and calculated molar volumes for heavy water. The calculation is slightly below the experiment at high temperature, but within the data scatter at low temperature. Indeed, in the supercooled region, as noted in Ref.~\onlinecite{holten_thermodynamics_2012}, ``there are differences of up to 0.14\% between the data
sets, and it is not clear which is the best set.'' Moreover, the uncertainties are not provided in some of the sources, making it difficult to assess the uncertainties on $\Theta$ and $\Pi$. We tried repeating the procedure after excluding the data sets which reach the lowest temperatures, Refs.~\onlinecite{zheleznyi_density_1969} or~\onlinecite{rasmussen_clustering_1973}, or both. We obtained for $(\Theta,\Pi)$ the values $(1.029,1.023)$, $(1.031,1.026)$, and $(1.027,1.021)$, respectively.

The corresponding states analysis achieves a similar accuracy as the thermal offset approach\cite{vedamuthu_simple_1996} (Fig.~\ref{fig:Vmolshift}), except at the highest temperatures. Overall, the agreement is satisfactory, as the influence of the LLCP, if it exists, is expected to be stronger at low temperature.

\begin{figure}
\centering
\includegraphics[width=0.94\columnwidth]{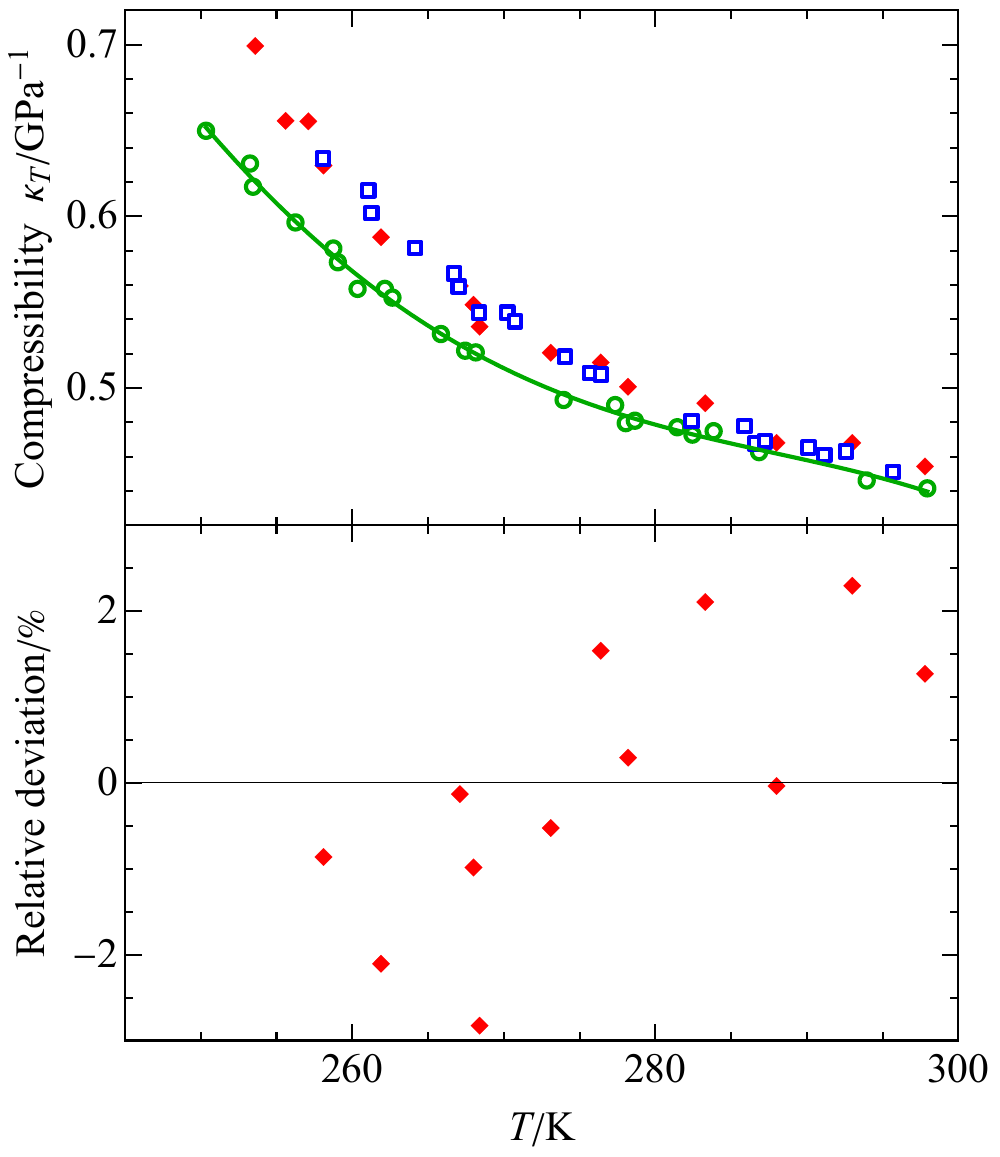}
\caption{Isothermal compressibility of water at \SI{10}{MPa}. Top panel: data for H$_2$O (empty green circles) with their fit (green curve), for D$_2$O (filled red diamonds), and results of Eq.~\ref{eq:rescalekT} applied to H$_2$O data (empty blue squares). Bottom panel: relative deviation of D$_2$O data from values calculated using Eq.~\ref{eq:rescalekT} combined with the fit to H$_2$O data. Here $\Theta=1.031$ and $\Pi=1.026$.\label{fig:kTCSA}}
\end{figure}

\textit{Isothermal compressibility.\label{sec:kT}} Isothermal compressibility $\kappa_T=-(1/V)(\partial V/\partial P)_T$ is central to the discussion of the putative LLCP in water, because $\kappa_T$ should diverge at this LLCP, and show maxima along isobars at pressures below $P_\mathrm{c}$. 

Robinson and collaborators considered only compressibility in H$_2$O~\cite{vedamuthu_properties_1995} but did not attempt to apply the thermal offset hypothesis. Kim~{\etal}~\cite{kim_temperatureindependent_2017} observed that $\kappa_T$ of D$_2$O at \SI{50}{MPa} could be superimposed on H$_2$O data after a \SI{6}{K} shift. Here we try the corresponding states analysis.  Writing $\kappa_\mathrm{H}$ and $\kappa_\mathrm{D}$ for light and heavy water, respectively, and assuming $\hat{P}\simeq 0$, Eq.~\ref{eq:CSA} predicts the following relation:
\begin{equation}
\kappa_\mathrm{D}(T) = \frac{1}{\Pi} \,\kappa_\mathrm{H}\!\left( \frac{T}{\Theta} \right)\, .\label{eq:rescalekT}
\end{equation}

To test Eq.~\ref{eq:rescalekT}, we use data from Kanno and Angell who measured $\kappa_T$ for both isotopes in the supercooled region at the same pressures~\cite{kanno_water_1979}. Figure~\ref{fig:kTCSA} shows an excellent agreement at low temperature for \SI{10}{MPa}. This is noteworthy as we did not use any further fitting parameters than $\Theta$ and $\Pi$ determined from the molar volumes. A good agreement is also observed at low temperature for other pressures (see the Appendix). As the pressure increases, the experimental D$_2$O data tend to be higher than the prediction with Eq.~\ref{eq:rescalekT}. We note however that, in principle, the comparison should be made at the same value of $P/P_\mathrm{c}$. For instance, when using the $\SI{100}{MPa}$ data for light water, heavy water data at $\SI{102.6}{MPa}$ should be used instead of $\SI{100}{MPa}$. From Ref.~\onlinecite{kanno_water_1979}, we estimate this would decrease the experimental values by $\simeq 1.5\%$, thus improving the agreement with the prediction.

\begin{figure}
\centering
\includegraphics[width=0.94\columnwidth]{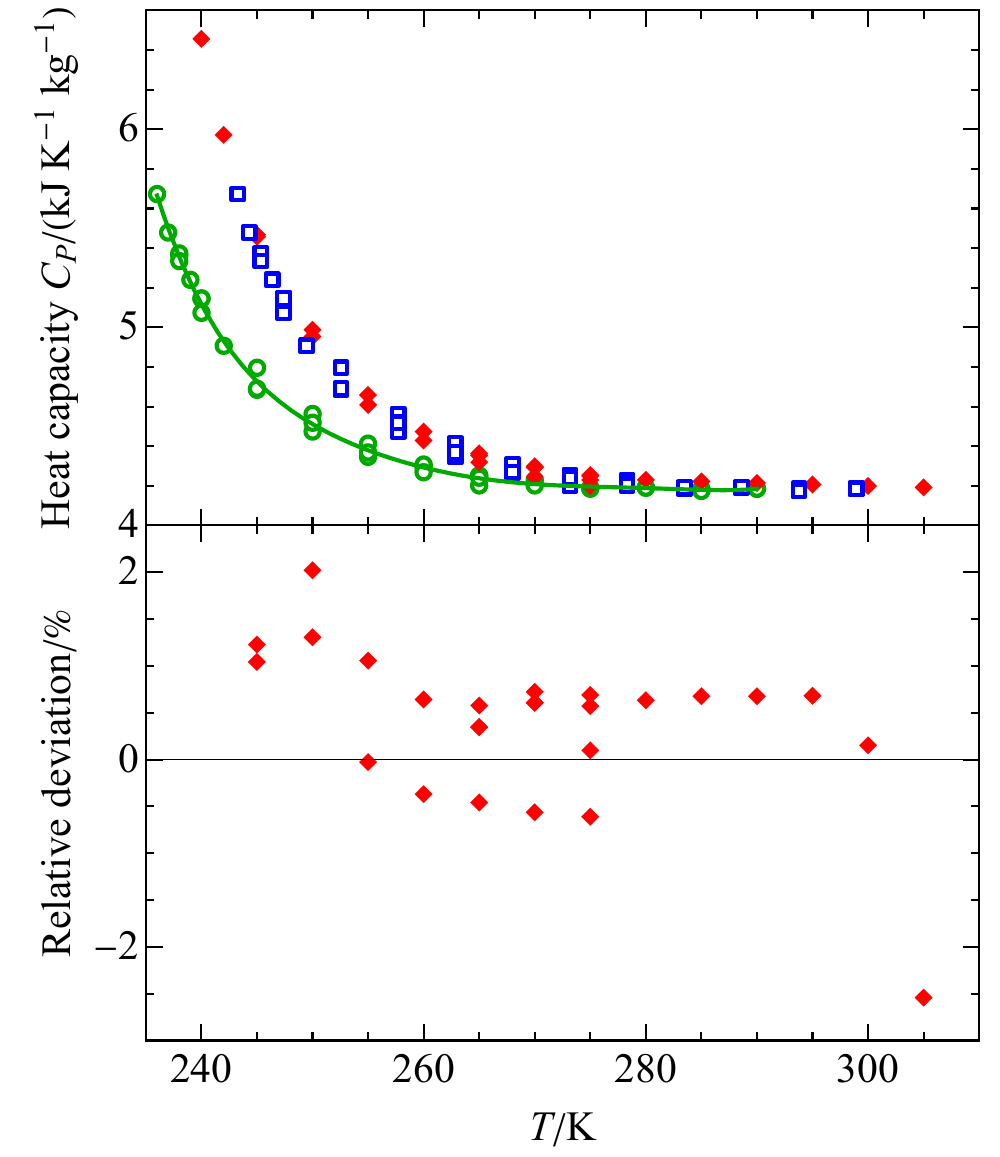}
\caption{Isobaric heat capacity of water. Top panel: data for H$_2$O (empty green circles) with their fit (green curve), for D$_2$O (filled red diamonds), and results of Eq.~\ref{eq:CPCSA} applied to H$_2$O data (empty blue squares). Bottom panel: relative deviation of D$_2$O data from values calculated using Eq.~\ref{eq:CPCSA} combined with the fit to H$_2$O data. Here $\Theta=1.031$.\label{fig:CPCSA}}
\end{figure}

\textit{Isobaric heat capacity.\label{sec:CPCSA}} We now consider the isobaric heat capacity $C_P$, and test the following relation:
\begin{equation}
C_{P,\mathrm{D}}(T) = C_{P,\mathrm{H}}\left(\frac{T}{\Theta}\right) \,.\label{eq:CPCSA}
\end{equation}
The only measurement on supercooled D$_2$O was performed at ambient pressure by Angell~\textit{et al.}~\cite{angell_heat_1982}. There are several measurements for supercooled H$_2$O, but they show some discrepancies; we used Ref.~\onlinecite{angell_heat_1982} for consistency because both isotopes were measured in the same setup, and also because this experiment achieved the largest supercooling. Figure~\ref{fig:CPCSA} shows that the increase in $C_P$ upon cooling starts at higher temperature for D$_2$O, but the two data sets are superimposed after multiplying the temperatures of the H$_2$O data by $\Theta$. This is actually surprising, because Fig.~\ref{fig:CPCSA} involves heat capacities per unit mass, rather than the molar heat capacities, which would differ by 11\%. We note that the principle of corresponding states does not strictly apply to heat capacity. Indeed, in the case of the liquid-vapor transition, Guggenheim pointed out in his seminal work~\cite{guggenheim_principle_1945} that the molar $C_P$ scales as expected for monatomic fluids, but only approximately for diatomic molecules, the difference from monatomic fluids exceeding the theoretical free rotor contribution. Guggenheim attributed this to ``a small restriction of the rotation increasing with decreasing temperature''. Water molecules do not rotate freely, and possess vibrational degrees of freedom; it seems that the various contributions to heat capacity per unit mass become identical for the two isotopes after the temperature is rescaled. Understanding the isotopic effect on $C_P$ will require further investigation.

\begin{figure}
\centering
\includegraphics[width=0.94\columnwidth]{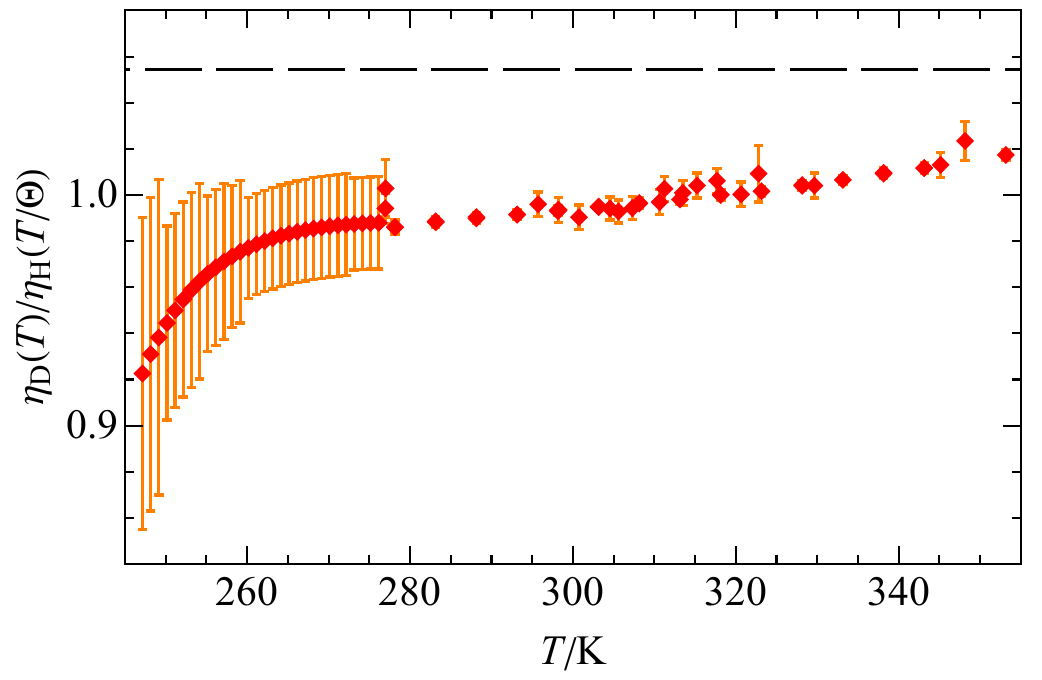}\\[1mm]
\includegraphics[width=0.94\columnwidth]{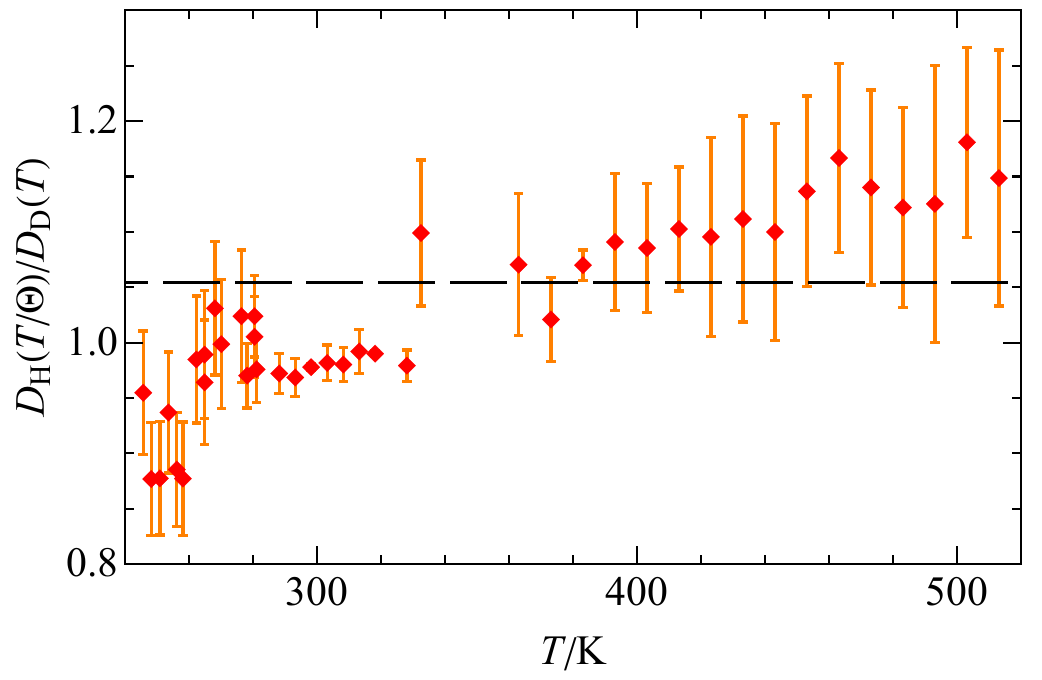}\\[1mm]
\includegraphics[width=0.94\columnwidth]{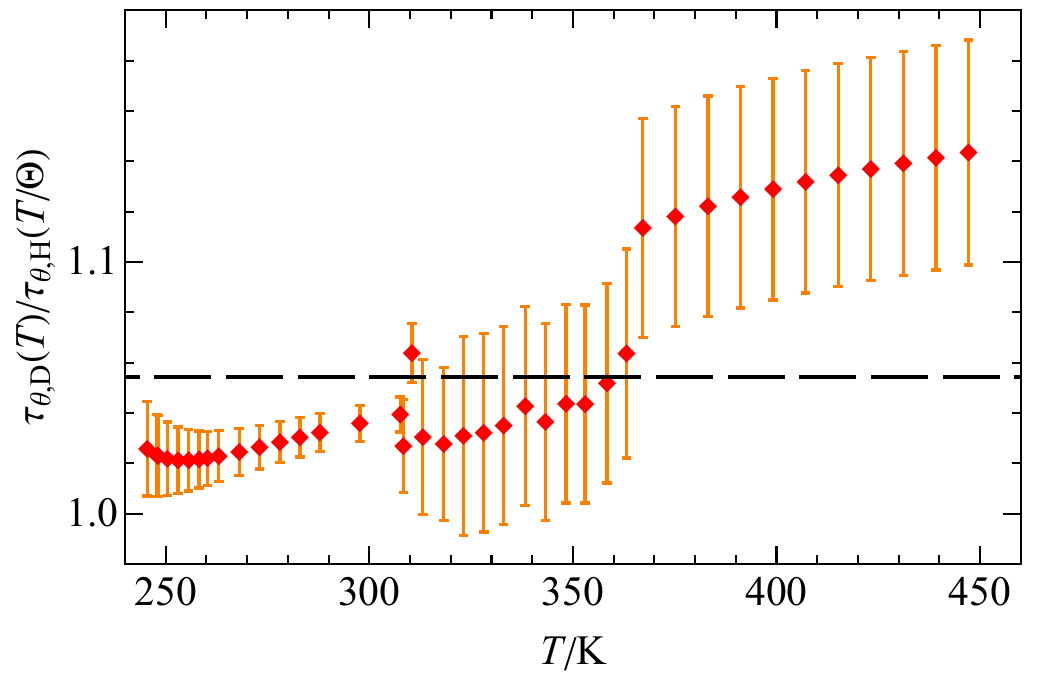}
\caption{Rescaled isotopic ratio $\widetilde{X}_\mathrm{r}$ for $X=\eta$ (top), $1/D$ (middle), and $\tau_{\theta}$ (bottom), using Eq.~\ref{eq:dynCSA} with $\Theta=1.031$. The error bars indicate one standard deviation. The horizontal dashed lines show $\sqrt{M_\mathrm{r}}$. \label{fig:dynCSA}}
\end{figure}

\textit{Dynamic properties.\label{sec:dyn}} We now turn to dynamic properties. In contrast to the thermal offset hypothesis, we do not attempt to achieve a perfect rescaling. Indeed, as discussed above, there is no compelling reason that the isotopic ratio for dynamic quantities in molecular liquids should exactly scale as $\sqrt{M_\mathrm{r}}$. Therefore, for a dynamic quantity $X$, we rather plot a \textit{rescaled} isotopic ratio after rescaling the temperature:
\begin{equation}
\widetilde{X}_\mathrm{r}=\frac{X_\mathrm{D}(T)}{X_\mathrm{H}(T/\Theta)}\,.\label{eq:dynCSA}
\end{equation}
Here the subscripts H and D stand for light and heavy water, respectively. Figure~\ref{fig:dynCSA} shows the results for $X=\eta$, $1/D$, and $\tau_{\theta}$, respectively. The H$_2$O data (Table~\ref{tab:H2O}) were fitted with Speedy-Angell laws:
\begin{equation}
X=X_0\left(\frac{T}{T_s}-1\right)^{-\gamma} \, ,
\label{eq:PL}
\end{equation}
for use in Eq.~\ref{eq:dynCSA}.

The raw isotopic ratios $\eta_\mathrm{r}$, $D_\mathrm{r}$, and $\tau_{\theta,\mathrm{r}}$ reach high values at low temperature: 2.38, 1.66, and 3.26, respectively (see Fig.~\ref{fig:IE}). After rescaling the temperature, we see that all ratios lie close to $\sqrt{M_\mathrm{r}}$, but systematic deviations are observed. Still, it is interesting to see that the amplitude of the dynamic isotope effect is much reduced after this temperature rescaling, making it similar to a number of other molecular liquids.

\textit{Violation of the Stokes-Einstein relation.\label{sec:SER}} We believe another approach is required to apply the corresponding states analysis to dynamic quantities. We propose to use the Stokes-Einstein relation (SER). For a Brownian sphere of radius $R$ in a liquid with shear viscosity $\eta$, the diffusion coefficient $D$ obeys the SER: $D=\kB T/(C \pi \eta R)$, where $\kB$ is the Boltzmann constant, and $C$ a coefficient ranging from $6$ to $4$ for no-slip to full-slip boundary conditions, respectively\cite{sutherland_measurement_1904,sutherland_dynamical_1905}. It follows that the Stokes-Einstein ratio (SE ratio), $D \eta/T$, is temperature-independent. Usually, for a molecule of a liquid, the constancy of $D\eta/T$ (using now the self-diffusion coefficient $D$), holds at high temperature but fails when the temperature decreases below around $1.3\,T_\mathrm{g}$, where $T_\mathrm{g}$ is the glass transition temperature~\cite{chang_heterogeneity_1997}. The violation in water already starts at ambient conditions, which corresponds to above $2\,T_\mathrm{g}$~\cite{dehaoui_viscosity_2015}.

\begin{figure}
\centering
\includegraphics[width=0.94\columnwidth]{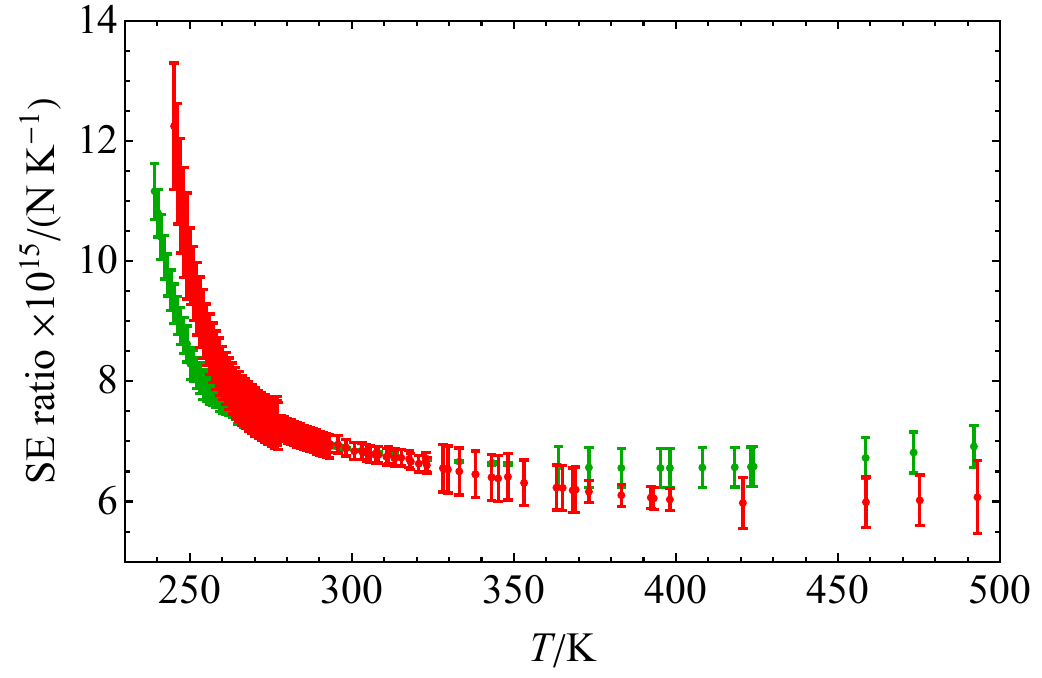}\\[1mm]
\includegraphics[width=0.94\columnwidth]{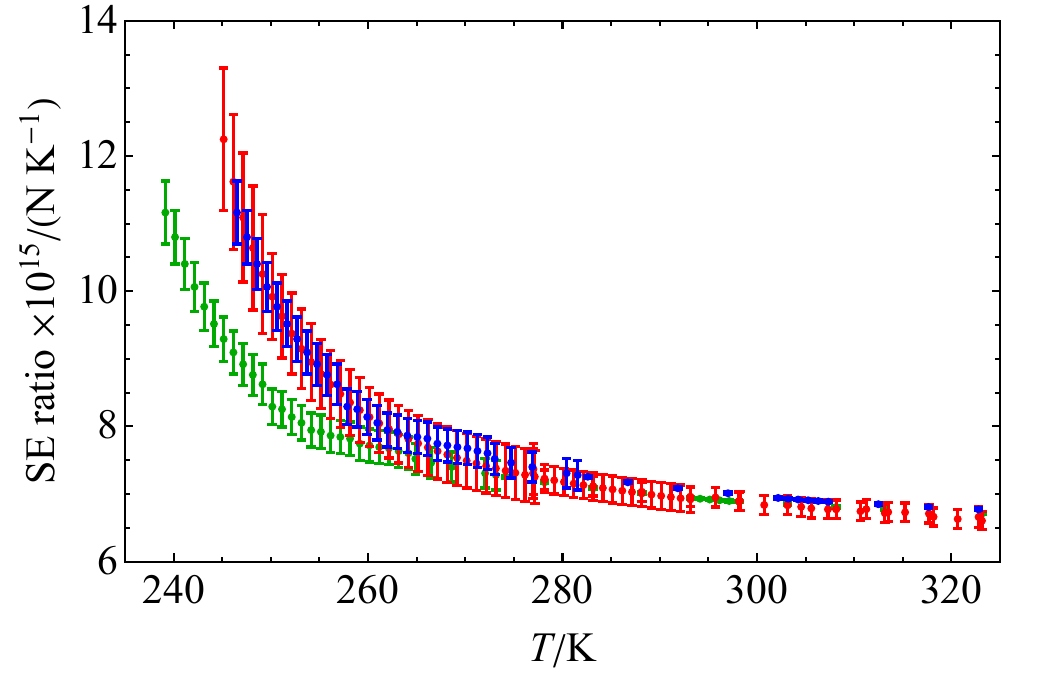}
\caption{Stokes-Einstein ratio $D \eta/T$ for light water (green) and heavy water (red) (top panel). The bottom panel shows a close-up at low temperature, adding light water data after multiplying their temperatures by $\Theta=1.031$ (blue). Error bars indicate one standard deviation.\label{fig:SECSA}}
\end{figure}

Our previous work on shear viscosity $\eta$ of supercooled light water~\cite{dehaoui_viscosity_2015}, combined with literature data on the self-diffusion coefficient~\cite{price_selfdiffusion_1999} and the rotational correlation time~\cite{qvist_rotational_2012}, revealed that the viscosity of water remains coupled with rotation in a fashion similar to usual glassformers, while it strongly decouples from translation. In particular, the SE ratio remains constant above room temperature but strongly increases upon cooling. Our recently obtained viscosity values in supercooled heavy water~\cite{ragueneau_shear_2022} confirmed a similar trend for the SER (see Fig.~\ref{fig:SECSA}, top panel). 

When the SER holds, we can compute an apparent hydrodynamic radius $\Rh$, defined by inverting the SER: $\Rh = \kB T/(C \pi D_\mathrm{s} \eta)$.
As shown in Ref.~\onlinecite{ragueneau_shear_2022}, above $280\,\mathrm{K}$, $\Rh$ is nearly the same for the two isotopes: $0.16$ or $0.11\,\mathrm{nm}$ with $C=4$ or 6, respectively. This common $\Rh$ value is also close to the size of a water molecule, which is virtually the same for H$_2$O and D$_2$O, as their molar volumes differ only marginally. The SER is thus fulfilled above $300\,\mathrm{K}$~\cite{ragueneau_shear_2022}. A close look at Fig.~\ref{fig:SECSA} reveals that the SER ratio above \SI{370}{K} is systematically slightly higher for H$_2$O than for D$_2$O. However, these H$_2$O values are based on $D$ from Ref.~\onlinecite{krynicki_pressure_1978}. Ref.~\onlinecite{yoshida_selfdiffusion_2008} provides another set of $D$ values in this temperature range, which are consistent within uncertainty with Ref.~\onlinecite{krynicki_pressure_1978}, but around 10\% lower. Using the lower values would reconcile the SER ratios for both isotopes at high temperature.

In contrast, at low temperature, SER is violated: the SE ratio increases sharply (see Fig.~\ref{fig:SECSA}), and the calculated $\Rh$ values become unphysically small, dropping by 40\% from 298 to $244\,\mathrm{K}$, and cannot be interpreted as a molecular size any more. The SER violation is usually attributed to decoupling between viscosity and translation, due to collective effects~\cite{dehaoui_viscosity_2015}. The distribution of relaxation times in the system, rather narrow at high temperature, broadens at low temperature and $\eta$ and $D$ decouple because they are related to different moments of this distribution~\cite{ediger_spatially_2000}. The fact that the two data sets in Fig.~\ref{fig:SECSA} run parallel to each other suggests that H$_2$O and D$_2$O experience similar collective effects, albeit starting at different temperatures. We may thus attempt the same corresponding states analysis, plotting the data for SE ratio in light water after multiplying their temperatures by $\Theta$. The result falls on top of the SE ratio in heavy water, see Fig.~\ref{fig:SECSA} (bottom panel). We argue that the direct comparison of the SE ratio $D\eta/T$ avoids the complications encountered with the prefactor introduced for the dynamic properties.

It would be interesting to repeat this analysis at higher pressure. We recently showed that the SE ratios for H$_2$O along various isobars are qualitatively similar. At room temperature, the apparent $R_\mathrm{h}$ values are nearly equal at all pressures, but the SER violation starts at lower temperature under pressure~\cite{mussa_viscosity_2023}. Unfortunately, along isobars at pressures above ambient, viscosity data for D$_2$O extends at most \SI{3}{K} below the melting point of each isobar~\cite{agayev_heavywater_1990}. Only the high temperature, rather flat part of the SE curves can thus be plotted for pressurized D$_2$O; within experimental uncertainty, they are compatible with both the raw and temperature-rescaled SE curves for H$_2$O. This calls for further viscosity measurements in supercooled D$_2$O under pressure.

\textbf{DISCUSSION\label{sec:disc}}

\begin{figure}
\centering
\includegraphics[width=0.94\columnwidth]{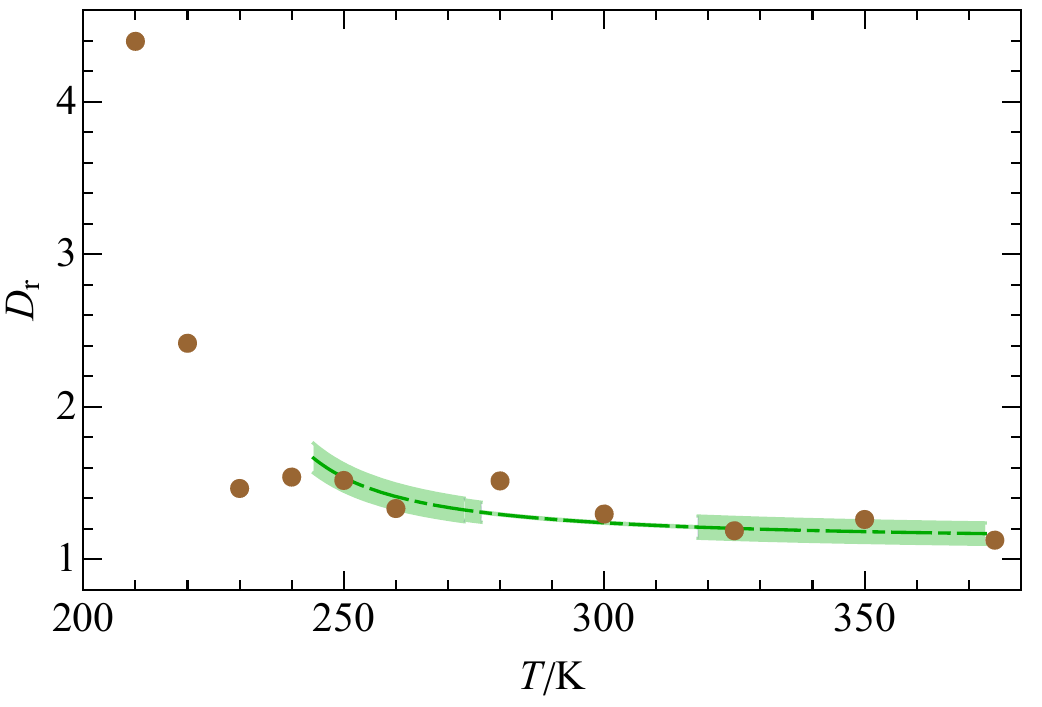}
\caption{Isotope effect as a function of temperature for self-diffusion $D_\mathrm{r}$ from the experiments (dash-dotted green curve) and path-integral molecular dynamics simulations~\cite{eltareb_nuclear_2021} (filled brown circles). The colored area denote $1-\sigma$ uncertainty.
\label{fig:IEPIMD}
}
\end{figure}

We have thus shown that molar volumes of light and heavy water can be superimposed in a corresponding states analysis, by choosing appropriate values of the parameters $\Theta=1.031$ and $\Pi=1.026$ in Eq.~\ref{eq:rescaleV}. Moreover, using fixed values for these parameters, a remarkable collapse is also achieved for isothermal compressibility and for the Stokes-Einstein ratio.

If a LLT does exist in supercooled water, the collapse of isothermal compressibility is expected as a consequence of the critical behavior dictating the divergence of $\kappa_T$ at the LLCP, and its reaching  maxima along the Widom line emanating from the LLCP~\cite{gallo_water_2016}. Actually, $\kappa_T$ maxima along isobars have been reported in H$_2$O at negative pressure~\cite{holten_compressibility_2017}, and in both H$_2$O and D$_2$O at near-zero pressure~\cite{kim_maxima_2017}. We can further test Eq.~\ref{eq:rescalekT} on this latter study, computing the ratios of temperatures and amplitudes for the respective maxima in D$_2$O (\SI{233.0\pm 1}{K}, \SI{97.5 \pm 0.5 E-5}{MPa^{-1}}) and H$_2$O (\SI{229.2\pm 1}{K}, \SI{104.5 \pm 1E-5}{MPa^{-1}}). The corresponding parameters would be $\Theta=1.017\pm 0.006$ and $\Pi=1.07 \pm 0.01$. This values are close to, but significantly different from, our values determined from the molar volumes. We note however that the exact location, and even existence, of the $\kappa_T$ maxima at $P=0$ is a matter of debate~\cite{caupin_comment_2018,kim_response_2018}.

A further comparison of $\Theta$ and $\Pi$ can be made using previous estimates of the LLCP coordinates for H$_2$O and D$_2$O. Interpreting his classic experiments on decompression- and compression-induced melting of high-pressure ices, Mishima proposed for the LLCP location ($T_\mathrm{c}\simeq \SI{220}{K}$, $P_\mathrm{c}\simeq \SI{100}{MPa}$)~\cite{mishima_decompression-induced_1998} and ($T_\mathrm{c} = \SI{230 \pm 5}{K}$, $P_\mathrm{c}\simeq \SI{50 \pm 20}{MPa}$)~\cite{mishima_liquidliquid_2000} for H$_2$O and D$_2$O, respectively. Later, analyzing volumes of supercooled H$_2$O, Mishima revised the LLCP location to ($T_\mathrm{c}\simeq \SI{223}{K}$, $P_\mathrm{c}\simeq \SI{50}{MPa}$)~\cite{mishima_volume_2010}. Taking Mishima's most recent estimates would give $\Theta \simeq 230/223 = 1.03$ and $\Pi \simeq 1$, in line with our values. In recent path-integral molecular dynamics simulations, Giovambattista and collaborators included from first principles nuclear quantum effects in the study of the phase diagram of water~\cite{eltareb_evidence_2022}. Their approach quantitatively captures the isotope effects on molar volumes and self-diffusion (see Fig.~\ref{fig:IEPIMD}). They reported LLT in both isotopes, with ($T_\mathrm{c}=\SI{159 \pm 6}{K}$, $P_\mathrm{c}= \SI{167 \pm 9}{MPa}$) and ($T_\mathrm{c} = \SI{177 \pm 3}{K}$, $P_\mathrm{c} = \SI{176 \pm 2}{MPa}$) for H$_2$O and D$_2$O, respectively. This yields $\Theta =1.11 \pm 0.05 $ and $\Pi=1.05 \pm 0.06 $, compatible with our values. We also note that both studies suggest a rather large value for $P_\mathrm{c}$, so that $P=\SI{0.1}{MPa}$ can be treated as $\hat{P}\simeq 0$.

We now turn to the origin of the collapse for the SE ratio. The violation of the SER in H$_2$O has been related to the putative liquid-liquid transition in supercooled water by molecular dynamics simulations~\cite{kumar_relation_2007}, phenomenologic two-state modelling~\cite{singh_pressure_2017}, or both~\cite{monterodehijes_viscosity_2018}. A recent molecular dynamics study~\cite{dueby_decoupling_2019} attributes the violation of SER upon cooling to the increasing role of molecular jumps in translational diffusion. When the jumps are removed, the residual diffusion due to cage trajectories of the molecules fulfills the SER. To explain the difference seen experimentally between isotopes, we propose that the share of jump trajectories in diffusion, due to structural changes in water, varies similarly for both isotopes, once the relative distance to the LLCP and the Widom line have been taken into account by the appropriate $T-P$ rescaling.

Structure factors of D$_2$O measured in the range $240-275\,\mathrm{K}$~\cite{kim_temperatureindependent_2017} were observed to match the structure factors of H$_2$O measured $\simeq\SI{5}{K}$ below. The same match can be obtained by rescaling the temperatures by a factor $\simeq 1.02$, close to $\Theta=1.031$. This gives credence to a structural explanation of the SE ratio collapse.

As just mentioned, a temperature shift closely resembles a temperature rescaling when a limited temperature range is considered. This can be easily seen, writing for a temperature $T_1 \simeq 260 \,\mathrm{K}$ the rescaled temperature $T_2 = \Theta T_1 = T_1 + (\Theta -1) T_1 \simeq T_1 + (\Theta -1) 260 \,\mathrm{K} = T_1 + 8 \,\mathrm{K}$. This shows that the comparable success of the two approaches, thermal offset and corresponding states analysis, is no coincidence. We argue that the latter approach is more physically grounded (being connected to the LLCP), gives consistent results for a single value of the parameter $\Theta$, and in addition gives an explanation for the amplitude rescaling involving the effect of pressure through $\Pi$. Nevertheless, Robinson's intuition is confirmed, and the role of zero-point and nuclear quantum effects is to move the LLCP location, as seen in PIMD simulations.

\begin{figure}
\centering
\includegraphics[width=0.94\columnwidth]{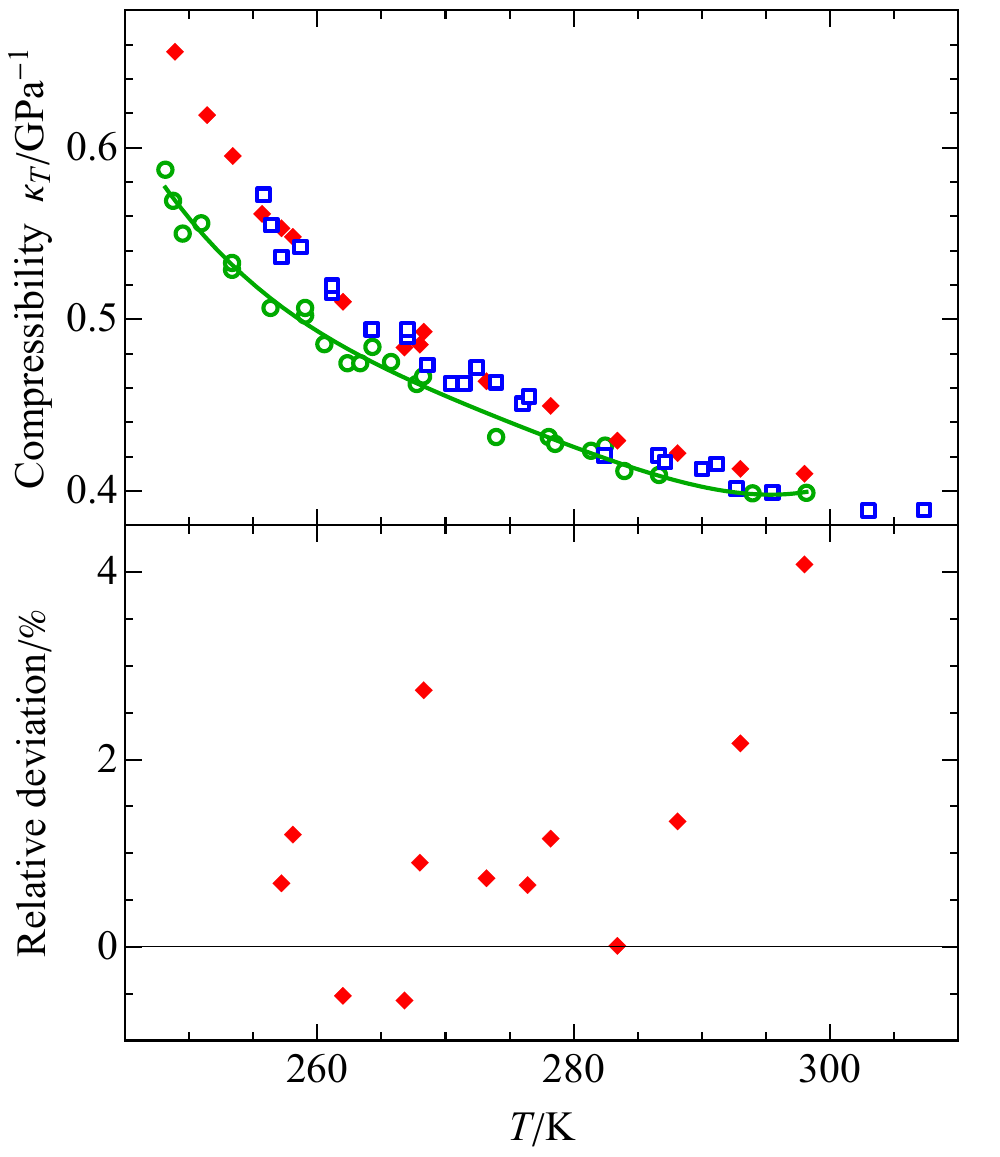}
\caption{Isothermal compressibility of water at \SI{50}{MPa}. Same legend as Fig.~\ref{fig:kTCSA}.\label{fig:kTCSAP50}}
\end{figure}

\begin{figure}
\centering
\includegraphics[width=0.94\columnwidth]{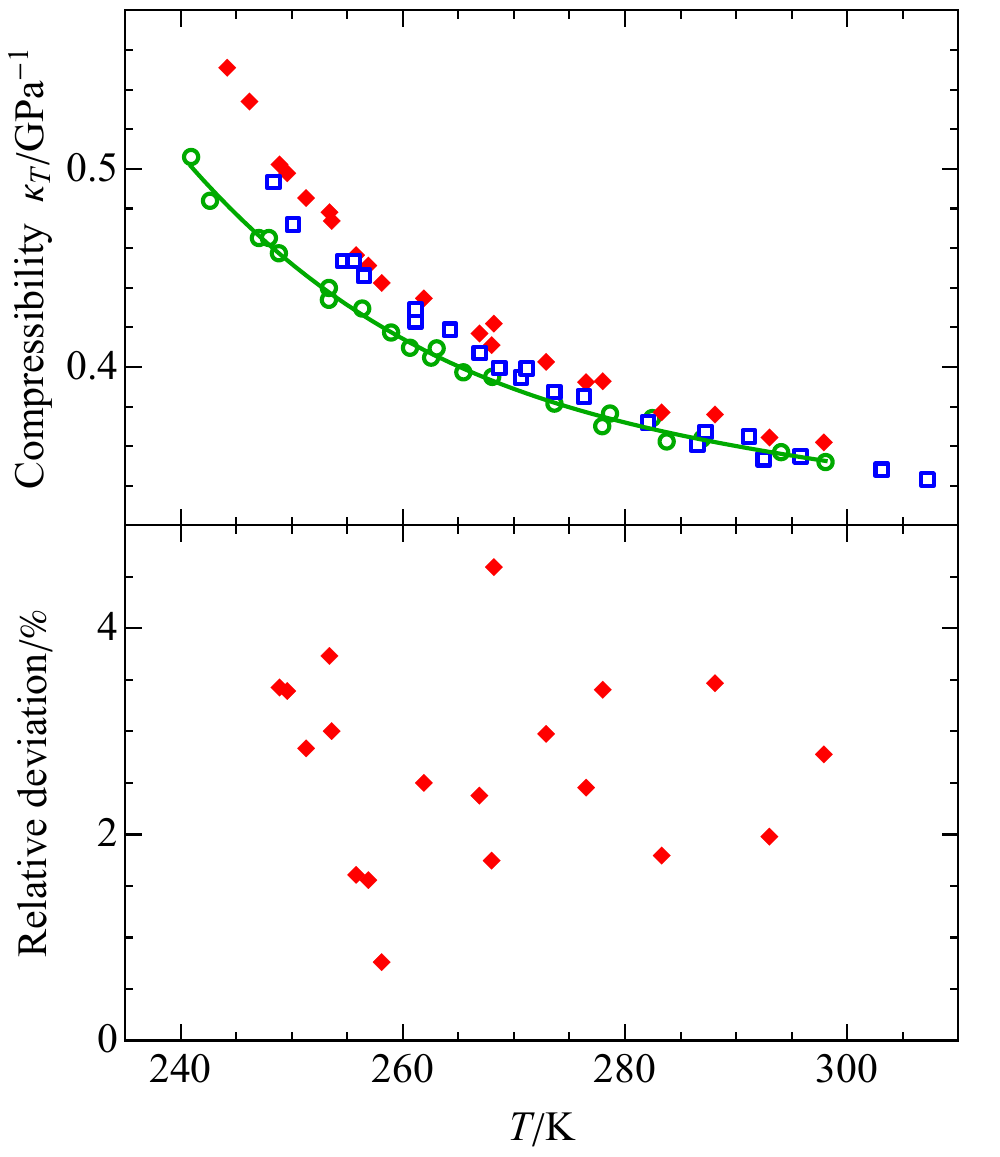}
\caption{Isothermal compressibility of water at \SI{100}{MPa}. Same legend as Fig.~\ref{fig:kTCSA}.\label{fig:kTCSAP100}}
\end{figure}

To conclude, the series of results presented here is thus consistent with the putative LLT in supercooled water, and our understanding of its manifestations. It does not however prove its existence and calls for further work.

\textbf{ACKNOWLEDGMENTS}

We thank Nicolas Giovambattista for providing the simulation data for $D_\mathrm{r}$, Gregory Kimmel for suggesting us to investigate the ``corresponding states'' approach, and Mikhaïl A. Anisimov for helpful discussions. We acknowledge support from Agence Nationale de la Recherche, grant number ANR-19-CE30-0035-01.

\appendix*
\textbf{APPENDIX: Isothermal compressibility under pressure}

Figures~\ref{fig:kTCSAP50} and~\ref{fig:kTCSAP100} show the results of the corresponding states analysis applied to isothermal compressibility data under pressure~\cite{kanno_water_1979}.

% Create the reference section using BibTeX:
%aipnum4-2.bst 2019-01-14 (MD) hand-edited version of apsrev4-1.bst
%Control: key (0)
%Control: author (8) initials jnrlst
%Control: editor formatted (1) identically to author
%Control: production of article title (0) allowed
%Control: page (1) range
%Control: year (1) truncated
%Control: production of eprint (0) enabled
%

\end{document}